\documentclass[aps,floatfix,showpacs, nofootinbib]{revtex4-1}
\usepackage{amsmath}
\usepackage{amsfonts}
\usepackage{amssymb}
\usepackage{graphics,epsfig}
\usepackage{graphicx}
\begin{document}
\title{More on McVittie's Legacy: A Schwarzschild - de Sitter black and white hole embedded in an asymptotically $\Lambda$CDM cosmology}
\author{Kayll Lake \cite{email} and Majd Abdelqader \cite{eemail}}
\affiliation{Department of Physics, Queen's University, Kingston,
Ontario, Canada, K7L 3N6 }
\date{\today}

\begin{abstract}
Recently Kaloper, Kleban and Martin reexamined the McVittie solution and argued, contrary to a very widely held belief, that the solution contains a black hole in an expanding universe. Here we corroborate their main conclusion but go on to examine, in some detail, a specific solution that asymptotes to the $\Lambda$CDM cosmology. We show that part of the boundary of the solution contains the inner bifurcation two - sphere of the Schwarzschild - de Sitter spacetime and so both the black and white hole horizons together form a partial boundary of this McVittie solution. We go on to show that the null and weak energy conditions are satisfied and that the dominant energy condition is satisfied almost everywhere in the solution. The solution is understood here by way of a systematic construction of a conformal diagram based on detailed numerical integrations of the null geodesic equations. We find that the McVittie solution admits a degenerate limit in which the bifurcation two - sphere disappears. For solutions with zero cosmological constant, we find no evidence for the development of a weak null singularity. Rather, we find that in this case there is either a black hole to the future of an initial singularity or a white hole to its past.
\end{abstract}
\pacs{04.20.Cv, 04.20.Ha, 98.80.Jk}
\maketitle

\section{Introduction}

A simple, but painful truth is the fact that it is far easier to find an exact solution to Einstein's
equations than it is to understand it. A fine example of this is given by McVittie's inhomogeneous cosmological solution \cite{mcvittie}, the
meaning of which has been debated since 1933. In retrospect,
this effort has to be considered an utterly remarkable
step into an area of research which is, to this day, still
in its infancy \cite{inhomogeneou}. The McVittie solution has been the
subject of a large number of investigations (we point to the recent thesis
by Martin \cite{martin} and the review given in \cite{car}), and generalizations \cite{car1}, but only recently did Kaloper, Kleban and Martin \cite{kkm} (henceforth KKM) explain the misinformation which has developed around this solution. In this paper we corroborate the main conclusion in KKM but also penetrate more deeply into an understanding of a specific solution. By way of the specification of a characteristic function for the solution, we exhibit a specific solution that asymptotes to the standard $\Lambda$CDM universe. Remarkably, part of the inner boundary of this solution contains the inner bifurcation two - sphere of the non - degenerate Schwarzschild - de Sitter spacetime. This tells us that both the black and white hole horizons of the extended Schwarzschild - de Sitter spacetime form part of the boundary of this McVittie solution, a possibility not envisioned in the KKM analysis. The specification of a characteristic function provides sufficient detail to allow for an examination of energy conditions and a systematic construction of the conformal diagram based on detailed numerical integrations of the null geodesic equations. The present work suggests that the very notion of an inhomogeneity in cosmology may well go beyond the concept of inhomogeneity in elementary physical variables. Finally, a minor point in the KKM analysis was the suggestion that in the case of a zero cosmological constant, the would - be horizon forms a weak null singularity. Here we find no evidence for this behaviour.\footnote{Whereas the analysis given here parallels, in some respects, that given in KKM, it also differs in a number of important aspects. We present a full discussion and point out at various stages agreement and disagreement with the KKM analysis. }

\section{The Solution}
\subsection{Overview}
The particular solution we are concerned with here is
the simplest of the McVittie class,\footnote{Recently, study of the McVittie solution has been denigrated \cite{krasinsky} with the view, in part, that this class of solutions is but a simple subset of a larger class of known solutions. The results presented in this paper argue for the contrary view. Even a simple looking metric can, when properly studied, yield a rich geometric structure.} and this can be written
in the form (e.g. \cite{nolan}) \cite{notation}
\begin{equation}\label{metric1}
    ds^2=-\left(\frac{1-m/2u}{1+m/2u}\right)^2dt^2+e^{\beta(t)}(1+m/2u)^4(dr^2+r^2d\Omega^2_{2})
\end{equation}
where $u\equiv r e^{\beta/2}$, $m$  is a positive constant and $d\Omega^2_{2}$ is
the metric of a unit 2-sphere. Clearly, for $m =0$ we obtain a spatially flat Robertson -
Walker metric, and for constant $\beta$, we have the Schwarzschild
metric (here in isotropic coordinates). These observations do
not constitute an understanding of the metric (\ref{metric1}). Indeed, it is a trivial exercise to construct distinct spacetimes with these two fundamental features.  Perhaps, the enduring interest in the McVittie solution derives from the observation that if we take the
coordinates of (\ref{metric1}) as comoving then we obtain a perfect
fluid with energy density $\rho$ and isotropic pressure $p$ given
by (e.g. \cite{nolan})
\begin{equation}\label{rho}
    8 \pi \rho = \frac{3}{4} \dot{\beta}^2, \;\; 8 \pi p = -\frac{3}{4} \dot{\beta}^2 - \frac{\ddot{\beta}}{\sqrt{1-2m/R(t,r)}}
\end{equation}
where $^{.} \equiv d/dt$ and
\begin{equation}\label{Rdef}
    R \equiv u(1+m/2u)^2.
\end{equation}
The uniform nature of the energy density and non-uniform nature of the pressure is often brought forward as a reason to consider this solution unphysical. However, even in the static Schwarzschild interior solution, such conclusions can be considered hasty \cite{mtw}. Our purpose here is not to argue, ab initio, for the physicality of the McVittie solution, but rather our purpose is to exhibit in detail the rather remarkable, and heretofore unrecognized, geometric structure that this solution presents. We comment on the idealization that the solution represents only once this structure is developed.

\bigskip

Under the coordinate transformation defined by (\ref{Rdef}), the metric (\ref{metric1}) becomes (e.g. \cite{nolan})
\begin{equation}\label{metric2}
    ds^2 = -f(t,R)dt^2-\frac{2H(t)R}{\sqrt{1-2m/R}}dtdR+\frac{dR^2}{1-2m/R}+R^2d\Omega^2_{2}
\end{equation}
where
\begin{equation}\label{fdef}
    f \equiv 1-2m/R-H^2R^2
\end{equation}
and $H$ is the Hubble function, given by $H = \dot{\beta}/2=\dot{a}/a$ where $a(t)$ is the usual scale factor for $m=0$. The
form (\ref{metric2}) is the basis for much of the analysis in KKM. Here we use the form (\ref{metric2}) and additional transformations suitable for numerical integrations. Let us note that the effective gravitational mass \cite{mass} associated with (\ref{metric2}) is not $m$, but rather $M$, given by
\begin{equation}\label{mass}
    M(t,R)=m+\frac{1}{2}H^2R^3.
\end{equation}

\subsection{The Function $H$}

Here we are not interested in arbitrary functions $H$,
but only those that reflect, in a general way, the background
cosmological model as it is currently understood.
In particular, we take $\dot{H }< 0$ for finite $t$,
\begin{equation}\label{Hcon}
    \lim_{t \rightarrow \infty}H=H_{0}>0,\;\;\lim_{t \rightarrow \infty}\dot{H}=\lim_{t \rightarrow \infty}\ddot{H}=0,\;\;\lim_{t \rightarrow 0}H= \infty,
\end{equation}
and
\begin{equation}\label{betalim}
     \lim_{t \rightarrow 0}\beta=-\infty.
\end{equation}
These general properties are in fact crucial to the present analysis.

\bigskip

As recognized in KKM, $t=0$ is not, in general, part of the spacetime.
To see this here, from the definition of $u$ and (\ref{betalim}), we find
$\lim_{t \rightarrow 0}u=0$ for all finite $r$. From the definition (\ref{Rdef}) then
we have
\begin{equation}\label{Rlim}
    \lim_{t \rightarrow 0}R=
\begin{cases}
 0 & \text{if }m =0 \\
 \infty & \text{if }m \neq 0\\
\end{cases}
\end{equation}
and so $t=0$ is not, for $m \neq 0$, part of the spacetime (and no limit
$m \rightarrow 0$ exists). This result depends on (\ref{betalim}) and relaxation of (\ref{betalim}) gives rise to other more involved possibilities not discussed here.

\subsection{Scalar Singularities}

Let us consider the singularities of (\ref{metric2}), as
revealed by scalars polynomial in the Riemann tensor. It
turns out that we need only report the Ricci scalar $\mathcal{R}$,
which we find is given by
\begin{equation}\label{ricci}
    \mathcal{R}=12H^2+\frac{6\dot{H}}{\sqrt{1-2m/R}},
\end{equation}
in agreement with KKM, since all other invariants, derived
from (partial) derivatives of the metric tensor no higher
than 2, add no new information. For $0 < t < \infty$, there
is a singularity at $R=2m \;(u=m/2)$, which, as is clear
from (\ref{metric1}), is spacelike in agreement with KKM. The apparent
singularity at $t = 0$, over the range $2m < R < \infty$ is, as
explained above, not part of the spacetime (\ref{metric2}) due to (\ref{betalim}).

\subsection{Asymptotics}

With conditions (\ref{Hcon}) let us also consider the asymptotic limit $t \rightarrow \infty$, and in
particular the roots to $f_{0} = 1-2m/R-H_{0}^2R^2=0$.
There are three cases:
$27m^2H_{0}^2>1$, for which there are no positive roots,
$27m^2H_{0}^2=1$, for which there is one (coincident) positive
root $R=3m$, and $27m^2H_{0}^2<1$ for which there are
two distinct positive roots that satisfy $0 <2m < R_{-} < 3m < R_{+}$. It is only
the last case which is of central interest here. As explained in detail below, we are interested in solutions that asymptote to de Sitter space for $R$ and $t \rightarrow \infty$ and become Schwarzschild - de Sitter space for $R \rightarrow R_{-}$ and $t \rightarrow \infty$. This requires (e.g. \cite{lr})
\begin{equation}\label{h0lambda}
    H_{0}^2=\Lambda/3
\end{equation}
where $\Lambda$ is the cosmological constant. For $27m^2H_{0}^2 \ll 1$ we note that $R_{-} \simeq 2m$ and $R_{+} \simeq 1/H_{0}$, the Hubble length (e.g. \cite{jhm}). The solution discussed here is essentially homogeneous at this length scale.

\subsection{The Locus $f=0$}

\begin{figure}[ht]
\epsfig{file=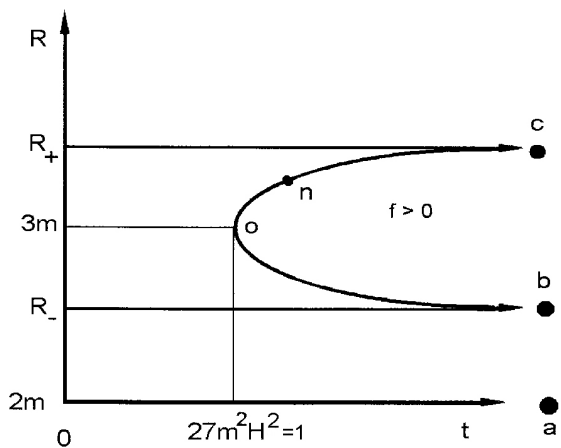,height=2.5in,width=3in,angle=0}
\caption{\label{figure1} The $R - t$ plane and the locus $f = 0$. The locus is
timelike above and to the right of $n$ where $\dot{H} = 2H(3m-R)/R^2$. The locus is spacelike
below $n$. $R = 3m$ is also shown. It passes through $f = 0$
at $o$ where $t = T$ such that $27m^2H(T)^2 = 1$. $R = 3M$
is spacelike to the left the locus and timelike to the right.
The trajectories $R_{\pm}$ are spacelike. The asymptotic points are
defined by $a: (t \rightarrow \infty, R = 2m$), b: ($t \rightarrow \infty, R = R_{-} )$ and
c: ($t \rightarrow \infty, R = R_{+}$). These are explained below.}
\end{figure}

Tangents to the
locus $f=0$, where $f$ is defined by (\ref{fdef}), are null for $\dot{R} = 2H^2R^2$, that is, $\dot{H} = 2H(3m-R)/R^2$, and timelike for $\dot{R}<2H^2R^2$. The tangents are
spacelike for $\dot{R} > 2H^2R^2$ and for $\dot{R} < 2H^2R^2$ along
the branch $dR < 0, dt > 0$. The essential features are
summarized schematically in Figure \ref{figure1}.

\section{Null Geodesics - Qualitative}

\subsection{Outgoing and Ingoing Geodesics}

Let us first examine, qualitatively, general properties of
the radial null geodesics ({\large $\eta$}) of (\ref{metric2}). These must satisfy
\begin{equation}\label{geodesic}
    \frac{dR}{dt}\bigg|_{\eta}=\sqrt{1-2m/R}\left(HR \pm \sqrt{1-2m/R}\right)
\end{equation}
in agreement with KKM. Now, whereas for the ``+" (``outgoing") branch
clearly $dR/dt > 0$, the ``-" (``ingoing") branch requires further examination.
From (\ref{fdef}) and (\ref{geodesic}) it follows that for the ``-"
branch $dR/dt > 0$ for $f < 0$, $dR/dt < 0$ for $f > 0$,
and $dR/dt = 0$ for $f = 0$. In particular, note that for
$f < 0$, $dR/dt > 0$ along both branches of (\ref{geodesic}). It is
appropriate here to note from (\ref{metric2}) that tangents to surfaces of constant finite
$t$ are spacelike for $R > 2m$ (and so for finite $t$ we set the future orientation $dt/d\lambda|_{\eta}>0$ for affine $\lambda$ increasing to the future) and tangents to surfaces of constant $R$ are
spacelike for $f < 0$, null for $f = 0$ and timelike for $f > 0$.

\subsection{Expansions}

Letting $k^{\alpha}_{\pm}$ signify the 4-tangents to the radial null geodesics, the associated expansions $\theta_{\pm} \equiv \nabla_{\alpha}k^{\alpha}_{\pm}$ follow as
\begin{equation}\label{expansions}
    \theta_{\pm}=\frac{2}{R} \sqrt{1-\frac{2m}{R}} \left(HR \pm \sqrt{1-\frac{2m}{R}}\right)\frac{dt}{d \lambda}\bigg|_{_{\pm}},
\end{equation}
where evaluation along {\large $\eta$} is now understood. This is in agreement with KKM up to the last term which is missing in the KKM analysis. The change in sign of $\theta_{-}$ at $f=0$ led KKM to refer to $f=0$ as an ``apparent horizon". Normally, this description would be reserved for a change in sign of $\theta_{+}$ since a change in sign of $\theta_{-}$ does not hide events below $f=0$ from distant observers. Now to see the importance of the last term in (\ref{expansions}) consider $t \rightarrow \infty$ along ingoing geodesics. Clearly all but the last term $\rightarrow 0$. However, as we show below, $t \rightarrow \infty$ at finite $\lambda$ and so the last term in (\ref{expansions}) diverges and as a result, $\theta_{-}$ becomes indeterminate in these coordinates. The removal of this ambiguity is discussed below.

\subsection{Infinity}

Let us examine the ``outer" boundary of (\ref{metric2}): ($R \rightarrow \infty, \;t \rightarrow \infty$) $ \equiv \mathcal{I}^{+}$ (spacelike \cite{penrose}).
The radial null geodesic equations can be written in the form
\begin{equation}\label{affine}
    \frac{d^2R}{d \lambda^2}=R(\lambda)\sqrt{1-2m/R(\lambda)}\frac{dH}{d t}\left(\frac{dt}{d \lambda}\bigg|_{_{\pm}}\right)^2 < 0
\end{equation}
where the inequality holds for finite $t$. From (\ref{affine}) it follows that $R \nrightarrow \infty$ for finite $\lambda$ and so both branches are future null geodesically complete \footnote{The same conclusion has been obtained by Brien Nolan (private communication).}. This is no surprise since $\mathcal{I}^{+}$ is indistinguishable from $\mathcal{I}^{+}$ for de Sitter space
(given (\ref{h0lambda})).
The ``inner" boundary of (\ref{metric2}) requires a much more detailed analysis. This is explained below. Here we simply introduce the notation: ($R= R_{-} ,t = \infty$) $\equiv \mathcal{H}$ (null), and note that
\begin{equation}\label{hrelation}
    f|_{_{\mathcal{H}}}=(R-2M)|_{_{\mathcal{H}}}=0,
\end{equation}
and ($R_{-} < R <\infty, t = \infty$) $\equiv i^{+}$.

\section{A Specific form for $H$}

In order to examine the solutions to (\ref{geodesic}) we must first
specify a specific background function $H$, which satisfies the conditions (\ref{Hcon}) and (\ref{betalim}), since (\ref{geodesic}) appears to
have no analytic solution for general $H$. Here we choose
\begin{equation}\label{Hspecial}
    H = \frac{H_{0} \sinh(3 H_{0}t)}{\cosh(3H_{0}t)-1} = H_{0}\coth(\frac{3H_{0} t}{2})
\end{equation}
so as to reflect an asymptotic $\Lambda$CDM universe.\footnote{This choice does not limit all of what follows. For example, the completeness/incompleteness arguments given in Appendix A are unaffected by this choice.}

\section{Energy Conditions}

\begin{figure}[ht]
\epsfig{file=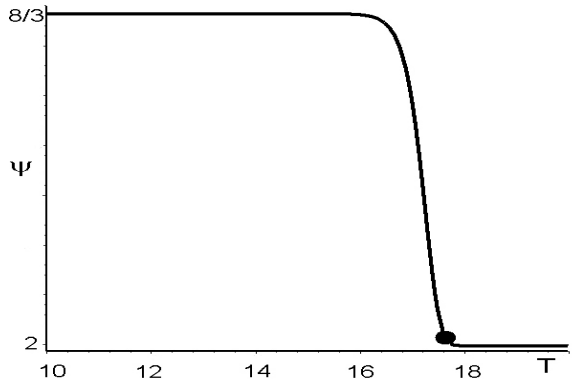,height=2in,width=3in,angle=0}
\caption{\label{figure2} The function $\psi$ given by (\ref{psispecial}). The curve has been constructed with $H_{0}=2.3\;10^{-18} s^{-1}$. $T$ is defined by $t=10^{T}$ where $[t] =s$. The current epoch is shown as a dot.}
\end{figure}

Note that we apply the classical energy
conditions to (\ref{metric1}) by way of the Einstein equations
(without an explicit cosmological constant). As is clear from (\ref{rho}), with our general conditions on $H$,
$\rho > 0$ and $\rho + p > 0$ and so the null and weak energy conditions
are satisfied. For the dominant energy condition,
$-\rho < p < \rho$ and so from (\ref{rho})
\begin{equation}\label{dominant}
    -\frac{3}{4} \dot{\beta}^2 < -\frac{3}{4} \dot{\beta}^2 - \frac{\ddot{\beta}}{\sqrt{1-2m/R}} < \frac{3}{4} \dot{\beta}^2.
\end{equation}
Whereas the left hand inequality is always satisfied, given our general conditions on $H$, the right hand side can be given in the form
\begin{equation}\label{psi}
    R(t) > \psi(t) m
\end{equation}
where, from (\ref{Hspecial}),
\begin{equation}\label{psispecial}
    \psi = \frac{2(\cosh(3H_{0}t)+1)^2}{\cosh(3H_{0}t)(\cosh(3H_{0}t)+2)}.
\end{equation}
Now whereas the explicit value of $\psi$ at any $t$ depends on $H_{0}$, the general form of $\psi$ does not. In particular, $\lim_{t \rightarrow 0}\psi = 8/3$ and $\lim_{t \rightarrow \infty}\psi = 2$. The function $\psi$ is shown in Figure \ref{figure2}. We conclude that the dominant energy condition is satisfied almost everywhere\footnote{We have no explanation for the curious position of the current epoch in Figure \ref{figure2}}.

Finally, the strong energy condition requires $\rho + 3 p \geq 0$. From (\ref{rho}) it follows that this condition requires
\begin{equation}\label{strong}
    \frac{2m}{R(t)} \geq 1 - \frac{\dot{H}^2}{H^4}.
\end{equation}
From (\ref{Hspecial}) we find
\begin{equation}\label{strongspecial}
   1-\frac{\dot{H}^2}{H^4} = \frac{(\cosh(3 H_{0}t)+4)(\cosh(3 H_{0}t)-2)}{(\cosh(3 H_{0}t)+4)^2}
\end{equation}
and so we arrive at
\begin{equation}\label{strongspecials}
         1-\frac{\dot{H}^2}{H^4} =
\begin{cases}
 < 0 & t < t_{0} \\
 = 0 & t = t_{0} \\
 > 0 & t > t_{0} \\
\end{cases}
\end{equation}
where
\begin{equation}\label{tnot}
   t_{0} = \frac{\ln(2 + \sqrt{3})}{2 H_{0}}
\end{equation}
and so the strong energy condition is satisfied everywhere for $t \leq t_{0}$. However, for $t > t_{0}$ the strong energy condition is satisfied only for
\begin{equation}\label{deltalim}
    R(t) < \delta(t)m
\end{equation}
where, clearly,
\begin{equation}\label{delta}
    \delta = \frac{2(\cosh(3H_{0}t)+4)^2}{(\cosh(3H_{0}t)+4)(\cosh(3H_{0}t)-2)}.
\end{equation}
Since, as it is easy to show, $\delta$ drops rapidly from $\infty$ at $t_{0}$ to $(\lim_{t \rightarrow \infty}\delta =) 2$, we conclude that the strong energy condition is eventually satisfied almost nowhere, as expected.

\section{Integration of the null geodesics}

\subsection{Integration of (\ref{geodesic}) in the $R - t$ plane}

\begin{figure}[ht]
\epsfig{file=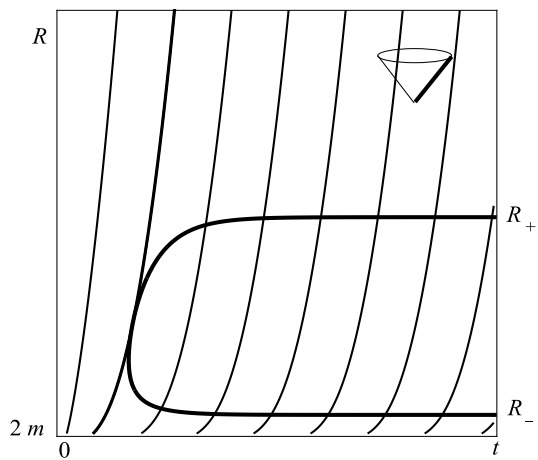,height=2.5in,width=3.5in,angle=0}
\caption{\label{figure3} The outgoing solutions of the null geodesic equation (\ref{geodesic}) under the condition (\ref{Hspecial}). The values taken are $H_{0} = 1/3$ and $m = (1/(H_{0}l_{0}))=958041/2000000 \sim 0.479$ (where $l_{0}$ is defined below). These values are of no consequence as we are simply interested in the qualitative behaviour of the solutions to (\ref{geodesic}). The locus $f = 0$ is also shown. The values of the roots $R_{\pm}$ are $R_{+} \sim 2.29$ and $R_{-} \sim 1.11$. The cone shows the ``leg'' of the null cone under consideration.}
\end{figure}

\begin{figure}[ht]
\epsfig{file=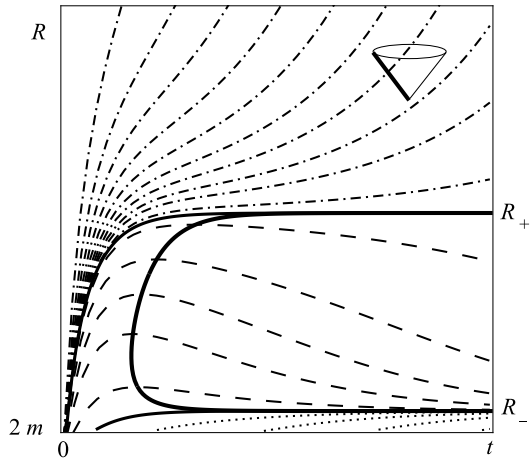,height=2.5in,width=3.5in,angle=0}
\caption{\label{figure4} The ingoing solutions of the null geodesic equation (\ref{geodesic}) under the same conditions as Figure \ref{figure3}.}
\end{figure}

We now examine numerical solutions to the null geodesic equations (\ref{geodesic}), subject to the choice (\ref{Hspecial}). First, consider the outgoing geodesics given by ``+" in (\ref{geodesic}). The integrations are shown in Figure \ref{figure3}. These geodesics are monotone in the $R - t$ plane. In contrast, the ingoing geodesics (``-" in (\ref{geodesic})) show considerably more structure. This is shown in Figure \ref{figure4}. We find 5 distinct types of evolution. Moving from the bottom right to the upper left in Figure \ref{figure4} we find: (i) Geodesics which asymptote monotonically to $\mathcal{H}$ (again, defined by ($R= R_{-} ,t = \infty$)) without crossing the locus $f=0$, (ii) A last geodesic that asymptotes monotonically to $\mathcal{H}$ without crossing $f=0$ (call it {\large $\eta$}$_{2}$), (iii) Geodesics which reach a maximum $R<R_{+}$ at $f=0$ and then asymptote monotonically to $\mathcal{H}$, (iv) A last geodesic that asymptotes monotonically to $R_{+}$ and terminates in $i^{+}$(call it {\large $\eta$}$_{1}$), and (v) Geodesics that cross $R_{-}$ and $R_{+}$ and monotonically evolve to $\mathcal{I}^{+}$. Now the cases (i) and the limit (ii), are absent in the KKM analysis, but are central to our examination of the inner boundary of (\ref{metric2}). Since the form (\ref{geodesic}) is particularly sensitive to error for these cases, we now introduce new coordinates to demonstrate, in more detail, that there do indeed exist ingoing null geodesics that reach $\mathcal{H}$ without crossing $f=0$.

\subsection{Integration of (\ref{geodesic}) in the $z - l$ plane}

We recast the problem as follows: Let \footnote{This useful definition for $z$ was pointed out to us by Brien Nolan (private communication).}
\begin{equation}\label{z}
    z \equiv \sqrt{1-\frac{2m}{R}},
\end{equation}
and let
\begin{equation}\label{l}
    l \equiv \frac{1}{Hm}.
\end{equation}
We observe the ranges
\begin{equation}\label{zrange}
    (R=2m)\;\;\; 0 < z < 1 \;\;\;(R \rightarrow \infty)
\end{equation}
and
\begin{equation}\label{lrange}
    (t \rightarrow 0)\;\;\; 0 < l < l_{0} \equiv \frac{1}{H_{0}m} \;\;\;(t  \rightarrow \infty).
\end{equation}

\begin{figure}[ht]
\epsfig{file=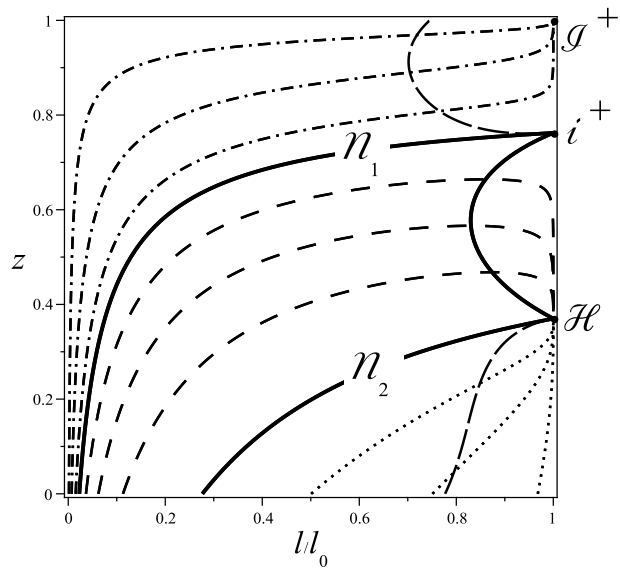,height=3in,width=3.5in,angle=0}
\caption{\label{figure5} Integration of (\ref{newgeodesic}) for $l_{0}=2000000/319347 \sim 6.263$ so that $z_{-}=37/100$ and $z_{+}=-37/200+35893^{1/2}/200 \sim 0.76$ over the range $0<l<l_{0}$. The limiting geodesics {\large $\eta$}$_{1}$ and {\large $\eta$}$_{2}$ are shown as is the locus $f=0$ (which connects $\mathcal{H}$ with $i^{+}$). The dashed curves connecting $z=0$ and $\mathcal{H}$ and $i^{+}$ with $z=1$ indicate $d^2z/dl^2=0$.}
\end{figure}

With the definitions (\ref{z}) and (\ref{l}), equation (\ref{geodesic}) takes the form (using the asymptotic $\Lambda$CDM model as before)
\begin{equation}\label{newgeodesic}
    \frac{dz}{dl} = \left( \frac{1-z^2}{l}-\frac{z(1-z^2)^2}{2}\right)\left(\frac{ l_{0}^2}{3(l_{0}^2-l^2)}\right).
\end{equation}
for the ingoing case. At first sight this might not appear to simplify things, but it does. First note that we need only specify $l_{0}$. For the case of central interest ($27 m^2H_{0}^2<1$) we have
\begin{equation}\label{lnot}
    l_{0} > 3 \sqrt{3}.
\end{equation}
As regards initial conditions, we have
\begin{equation}\label{initial}
    \frac{dz}{dl}\bigg|_{z=0,l=l_{1}} = \frac{l_{0}^2}{3 l_{1}(l_{0}^2-l_{1}^2)}
\end{equation}
which is regular over the range
\begin{equation}\label{lrange1}
    0 < l_{1} < l_{0}.
\end{equation}
Moreover, we have
\begin{equation}\label{Rdfdl}
    \frac{dz}{dl}
\begin{cases}
 >0 & f<0, l<\frac{2}{z(1-z^2)} \\
 =0 & f=0, l=\frac{2}{z(1-z^2)} \\
 <0 & f>0, l>\frac{2}{z(1-z^2)} .\\
\end{cases}
\end{equation}
Numerical integrations are shown in Figure \ref{figure5}. Our conclusion is that null geodesics that reach $\mathcal{H}$ without crossing $f=0$ are a fundamental feature of this solution.

\section{Global structure of the spacetime}

\subsection{Construction of the Conformal Diagram}
\begin{figure}[ht]
\epsfig{file=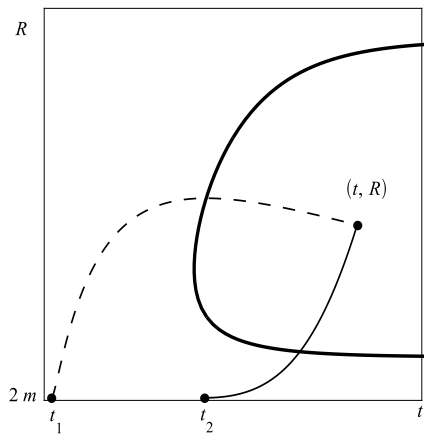,height=2.3in,width=3.5in,angle=0}
\caption{\label{figure6}Locating $t_{1}$ and $t_{2}$ for any event via null geodesics. The locus $f=0$ is shown. Note that $t$ increases to the right.}
\end{figure}

\begin{figure}[ht]
\epsfig{file=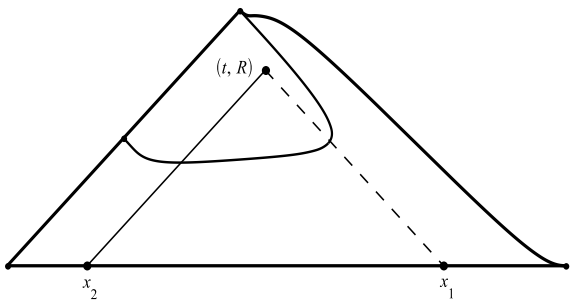,height=2.3in,width=3in,angle=0}
\caption{\label{figure7}Conformal representation of the event $(t,R)$. The locus $f=0$ is shown. Note that $t$ increases to the left.}
\end{figure}

The conformal representation of a spacetime (Penrose - Carter diagram \cite{griffiths}) is, of course, not unique in detail. However, all conformal representations must show the global structure of the spacetime.\footnote{All cases considered here are time-symmetric in the sense that the diagrams can be flipped upside down. Moreover, all diagrams can be rotated, interchanging the left and right hand sides.} Here we construct the Penrose - Carter diagram in the following way: The construction of the diagram starts by solving the null geodesic equations numerically. This gives a general understanding of the global behavior of the spacetime. In the present case, all of the outgoing geodesics and all of the ingoing geodesics originate from the singularity at $R=2m$. These geodesics intersect with $R=2m$ at some finite value of $t>0$. That is, any point in the spacetime ($t>0,R>2m$) can be connected to the past boundary ($R=2m, t>0$) by a unique null geodesic from each branch. We represent the boundary $R=2m$ as a horizontal line in a Cartesian plane ($y=0, -1\leq x \leq 1$), setting the right end of $R=2m$ at $t=0$, and the left end at $t=\infty$. To represent the interval $ 0<t<\infty $, from $x=-1$ to $1$, we use the transformation function
\begin{equation}\label{conformalT}
t=A(1-x)^B{\tan\left((1-x)\frac{\pi}{4}\right)},
\end{equation}
where $A$ and $B$ are adjustable constants. $A$ represents the value of $t$ at the center of the line $x=0$ and $B$ adjusts the position of $b$. These constants have no physical significance and where adjusted only for visual presentation (we chose $A \sim 0.39$ and $B \sim 0.75$). Once this line is adopted \cite{majd}, any point in the spacetime can be projected onto the conformal diagram by solving for both null geodesics that pass through any point, and numerically solving for the value of $t$ at which these two curves reach $R=2m$; say $t_1$, and $t_2$. This is illustrated in Figure \ref{figure6}. Once these two values of $t$ are found, we can represent them in the conformal diagram by solving for $x$ in (\ref{conformalT}). This gives us $x_1$ and $x_2$. Finally, we find the conformal representation of the original point by finding the intersection of the same two null geodesics, but presented in the conformal diagram as $y=-x+x_1$ for one geodesic, and $y=x-x_2$ for the other. This is illustrated in Figure \ref{figure7}. By our choice (\ref{conformalT}), all diagrams are strongly compactified in the region $R > R_{+}$.

\subsection{Null geodesics}

\begin{figure*}[ht]
\epsfig{file=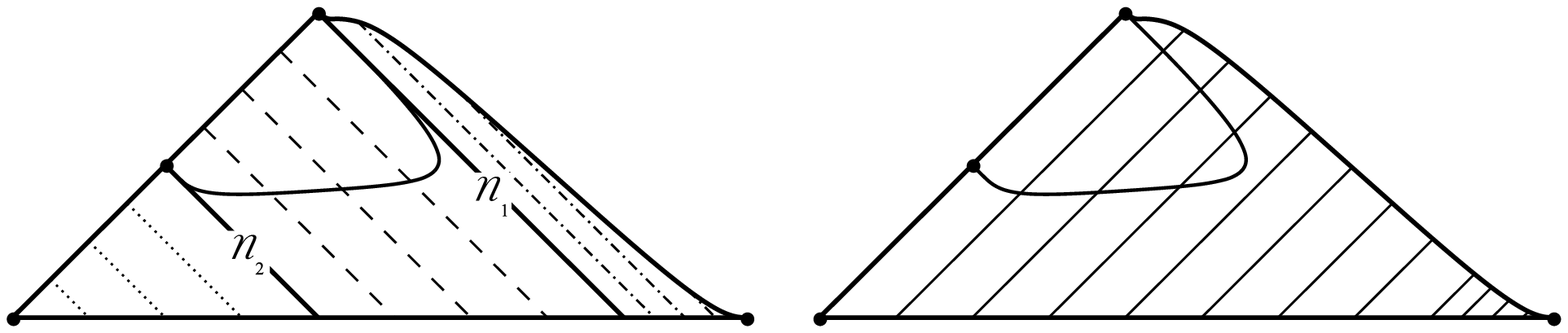,height=3.5in,width=6in,angle=0}
\caption{\label{figure8} At right, the conformal representation of the outgoing null geodesics as given in Figure \ref{figure3}. The locus $f = 0$ is shown. The bottom horizontal line represents the singularity $R=2m$. At left the conformal representation of the ingoing null geodesics as given in Figure \ref{figure4}.}
\end{figure*}

Under the procedure described above, Figure \ref{figure3} is mapped into the right hand side of Figure \ref{figure8} and Figure \ref{figure4} is mapped into the left hand side of Figure \ref{figure8}.

\subsection{Surfaces of constant $R$ and $t$}

\begin{figure*}[ht]
\epsfig{file=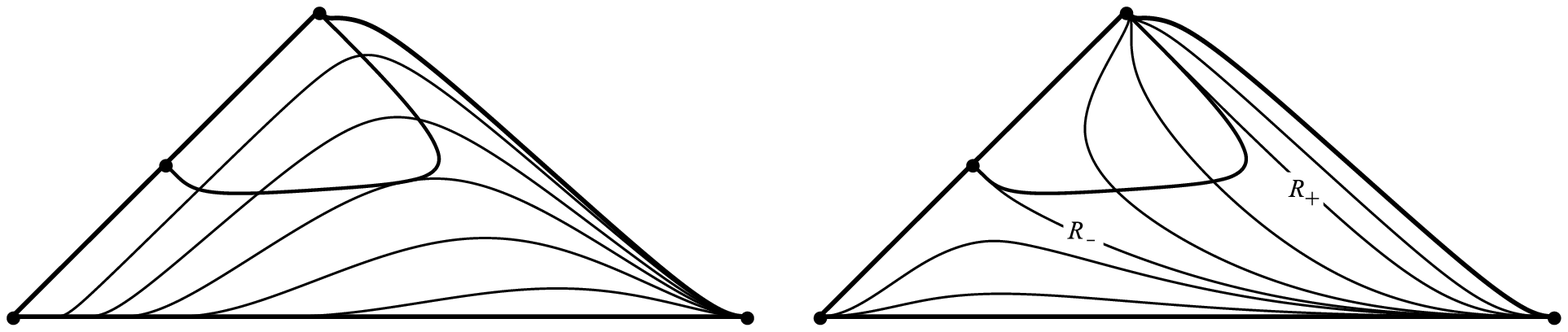,height=3.5in,width=6in,angle=0}
\caption{\label{figure9} At left the conformal representation of surfaces of constant $t$. The locus $f = 0$ is also shown. Moving from the bottom right to the upper left we have: $t_{0}<t < T$ (where $27m^2H(T)^2 = 1$), $t=T$, $t=t_{1} > T$ and $t_{2} > t_{1}$. The boundary has $t \rightarrow \infty$. At right the conformal representation of surfaces of constant $R$. The locus $f = 0$ is also shown. Moving from the bottom left to the upper right we have: $2m < R_{0}< R < R_{-}$, $R = R_{-}$, $R_{-} < R < 3m$, $R=3m$, $R=R_{+}$, $R_{1} > R_{+}$. The right boundary has $R \rightarrow \infty$ and the left boundary has $R = R_{-}$.}
\end{figure*}

Using the same procedure, at the left in Figure \ref{figure9} we show surfaces of constant $t$ and at the right in Figure \ref{figure9} we show surfaces of constant $R$. The surfaces of constant $t$ are spacelike for all finite $t$. The surfaces of constant $R$ are spacelike for $f<0$ and timelike for $f>0$.

\subsection{The fluid streamlines}

\begin{figure}[ht]
\epsfig{file=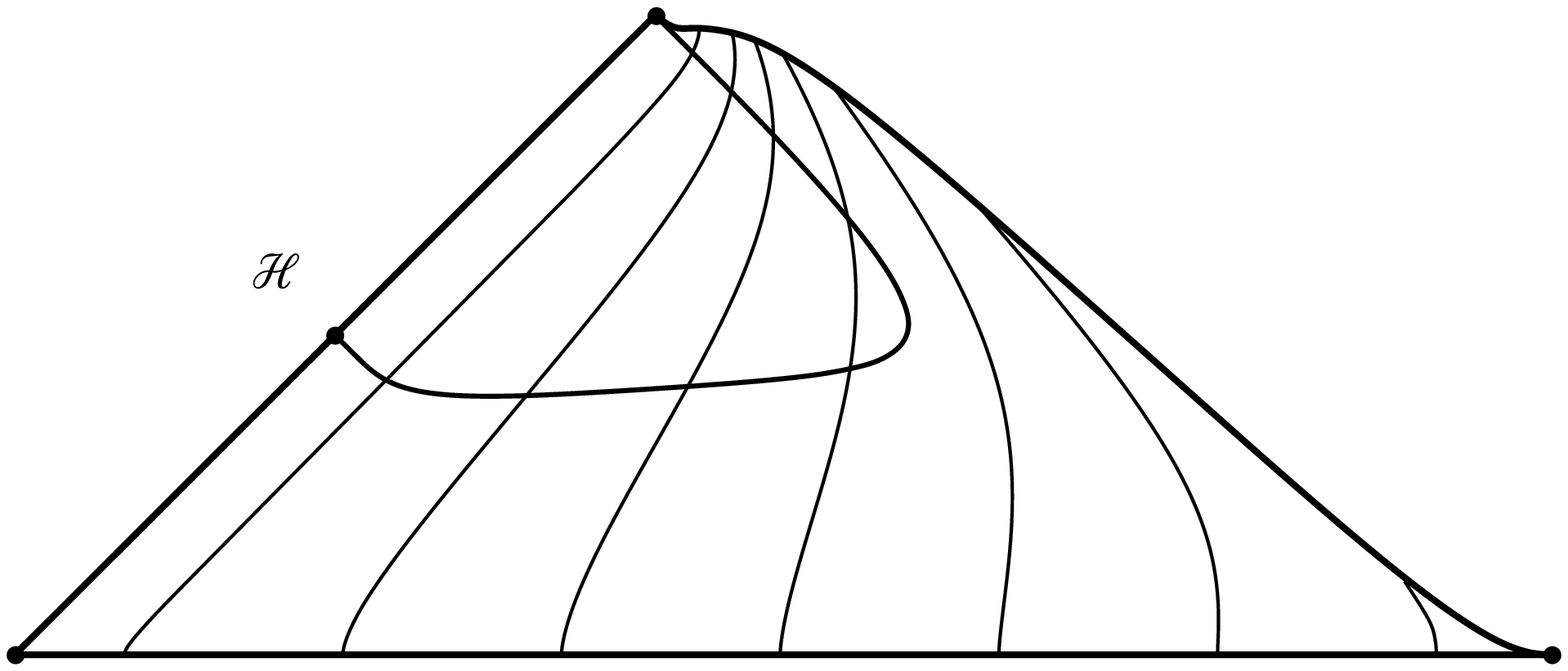,height=3in,width=3in,angle=0}
\caption{\label{figure10} The fluid streamlines $r = constant > 0$ in the conformal diagram. The streamlines are timelike and $r$ increases to the right. Note that the streamlines do not cross $\mathcal{H}$ which is the limit $r \rightarrow 0$.}
\end{figure}

The streamlines $r=constant$ can be written out explicitly in $z - l$ coordinates in the form
\begin{equation}\label{streamz}
 z(l) = \tanh \left(\frac{1}{6} \log \left( \left(\frac{l^2}{l_{1}^2}\right)\left(\frac{l_{0}^2-l_{1}^2}{l_{0}^2-l^2}\right)\right)\right)
\end{equation}
where $z(l_{1})=0$. Transforming to the $R - t$ plane, these streamlines take the form
\begin{equation}\label{streamr}
R(t)=2m \cosh \left(\frac{1}{6} \log \left(\frac{H(t_{1})^2-H_{0}^2}{H(t)^2-H_{0}^2}\right)\right)
\end{equation}
where $R(t_{1})=2m$. These streamlines are shown in Figure \ref{figure10}. Note that the streamlines do not cross $\mathcal{H}$. This fact can be considered the source of relation (\ref{hrelation}) and is the central part of the original construction \cite{mcvittie}.

\subsection{The boundary and ingoing geodesics}

\begin{figure}[ht]
\epsfig{file=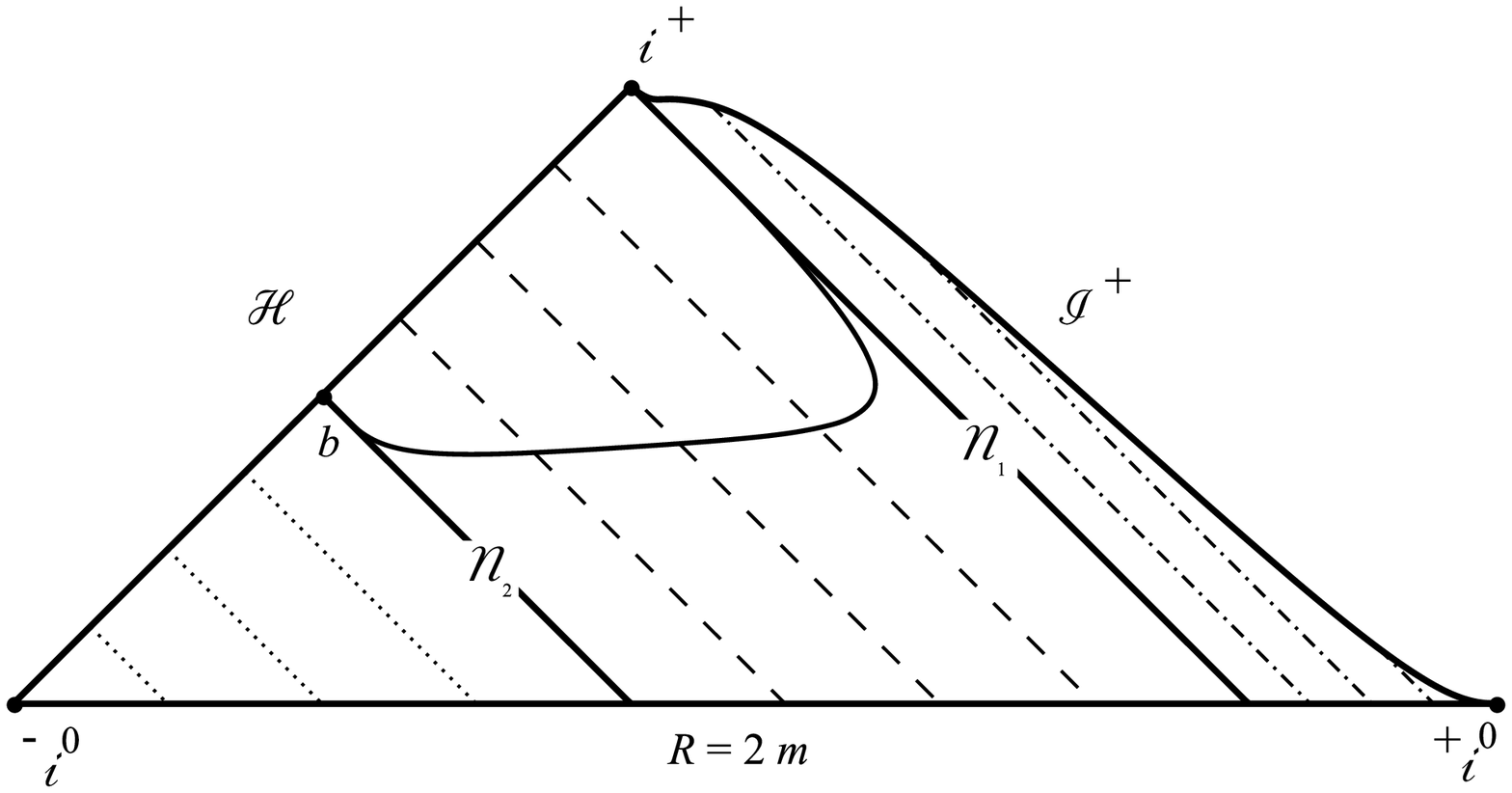,height=3in,width=3in,angle=0}
\caption{\label{figure11} The conformal representation of the boundary to the spacetime (\ref{metric2}) given (\ref{Hspecial}). The boundary is defined as follows: The singularity $R=2m$, $\mathcal{I}^{+} \equiv (R=\infty, t=\infty)$, $ i^{+} \equiv (R_{-} < R < \infty, t = \infty)$, $\mathcal{H} \equiv (R= R_{-} ,t = \infty)$, $^{-}i^{0} \equiv (2m \leq R < R_{-}, t \rightarrow \infty)$ and $^{+}i^{0} \equiv (2m \leq R < \infty, t >0)$.}
\end{figure}

In Figure \ref{figure11} we summarize the boundary of (\ref{metric2}) and classify ingoing null geodesics which do not terminate on $\mathcal{I}^{+}$. We are now in a position do discuss the affine completeness/incompleteness of these geodesics. As is discussed in detail in Appendix A, all these geodesics are incomplete except {\large $\eta$}$_{1}$ which we find to be complete. This incompleteness is the central point in KKM, but their analysis revealed only the section $(b,i^{+})$ of $\mathcal{H}$. Here we observe that {\large $\eta$}$_{2}$ terminates at $b$ on $\mathcal{H}$ and we observe very special properties associated with $b$ as explained in Appendix B: $b$ is characterized by vanishing expansion for both the ingoing and outgoing radial null geodesics. This is the hallmark of a bifurcation two - sphere. This bifurcation two - sphere divides $\mathcal{H}$ into two sections: a ``black hole" horizon to the future of $b$ and a ``white hole" horizon to the past of $b$. This is explored in the completion to the spacetime given below.

\subsection{A completion}

\begin{figure*}[ht]
\epsfig{file=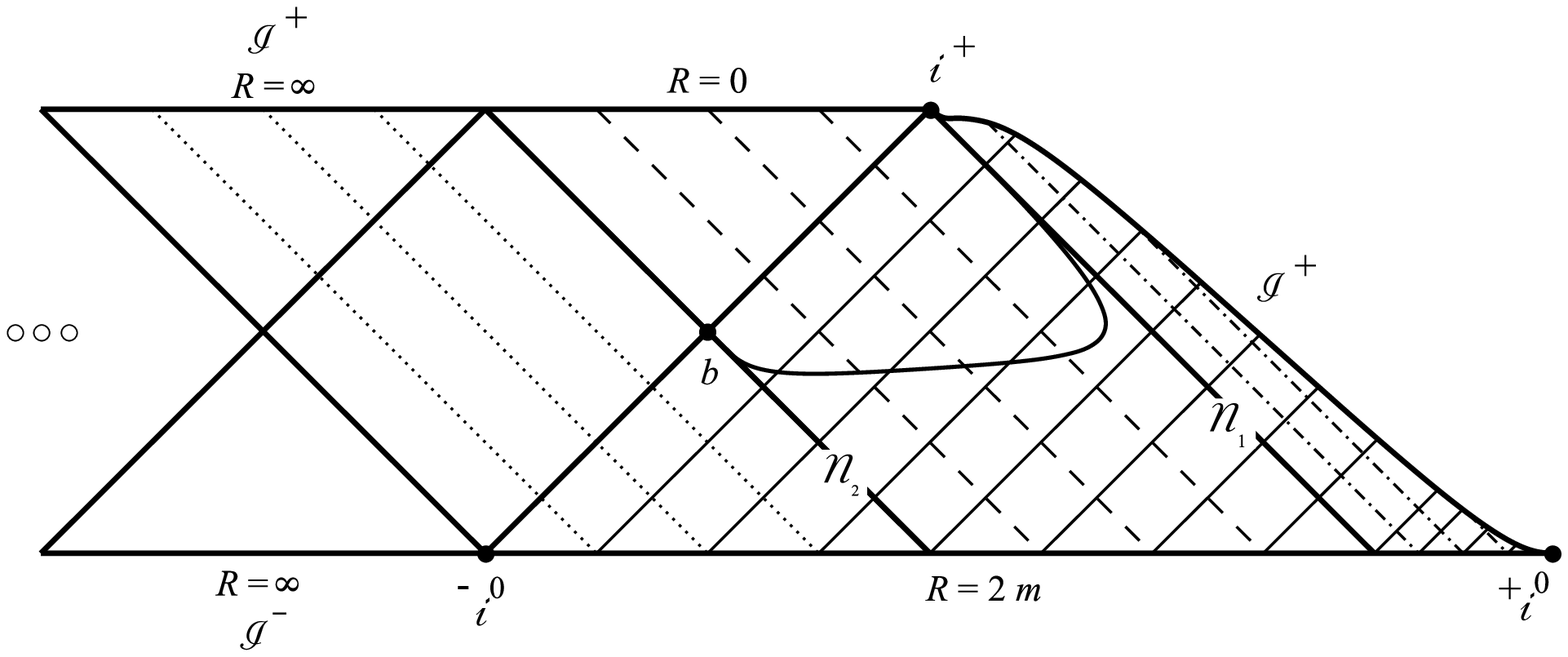,height=3.5in,width=6in,angle=0}
\caption{\label{figure12} An extension of the McVittie spacetime which is null geodesically complete. The inner bifurcate two - sphere of the Schwarzschild - de Sitter spacetime ($b$) is identified with that of the McVittie spacetime. }
\end{figure*}

One possible extension of the McVittie spacetime, which is null geodesically complete, is shown in Figure \ref{figure12} where we have identified the inner bifurcation two - sphere of the Schwarzschild - de Sitter spacetime with that of the McVittie spacetime \cite{schds}. This enlarged spacetime now has a center of symmetry, the center of the Schwarzschild - de Sitter black hole at $R=0$. Ingoing null geodesics below {\large $\eta$}$_{1}$ but above {\large $\eta$}$_{2}$ terminate at the singularity $R=0$ in the Schwarzschild - de Sitter black hole. Whereas {\large $\eta$}$_{2}$ joins onto the ``left" black hole horizon of the Schwarzschild - de Sitter spacetime, all ``ingoing" geodesics below {\large $\eta$}$_{2}$ have $R$ monotonically increasing, pass through the white hole horizon and terminate at $R = \infty$, that is, $\mathcal{I}^{+}$ of the Schwarzschild - de Sitter spacetime. Whereas the degenerate case $27m^2H_{0}^2=1$ is outside the cases of interest from a physical point of view, it offers a very instructive limit from a mathematical point of view since {\large $\eta$}$_{1}$ and {\large $\eta$}$_{2}$ then coincide and $b$ disappears. This is examined in Appendix C.

\section{Discussion}
We have performed a detailed study of a particular McVittie solution, by way of the specification of a characteristic function, that asymptotes to the standard $\Lambda$CDM cosmology and that contains an inner boundary that is a slice of the extended Schwarzschild - de Sitter spacetime. We have found that this inner boundary contains a bifurcation two - sphere where the expansion of both the ingoing and outgoing radial null geodesics vanishes. To the future of this bifurcation on this inner boundary we have found a black hole horizon at finite affine distance and therefore we corroborate the main conclusion in the work of Kaloper, Kleban and Martin. In addition, however, we have found a white hole horizon to the past of the bifurcation which is also at finite affine distance. In the degenerate limit of this particular solution the bifurcation two - sphere and black hole horizon disappear leaving only the white hole horizon, also at finite affine distance. We have shown that the null and weak energy conditions are satisfied and that the dominant energy condition is satisfied almost everywhere. The global structure of the solution has been constructed systematically based on detailed numerical integrations of the null geodesic equations. In the case $H_{0}=0$, the work of Kaloper, Kleban and Martin suggested that the would - be horizon forms a weak null singularity. In Appendix D we argue that this is not the case. Moreover, we argue that this case is rather less interesting than $H_{0}>0$ since the solutions can have a black hole horizon to the future of the singularity at $R=2m$ or a white hole horizon to the past of $R=2m$.

The present analysis relies on the (standard) condition (\ref{betalim}). Relaxing this, it is clear that we can maintain the conditions (\ref{Hcon}) but also violate (\ref{Rlim}). This means that in general the initial singularity will consist of two parts: the ``pressure" singularity at $R=2m$, and a generalized ``big bang" singularity at $t=0$. In a sense this shows how ``delicate" the solution is. The fact that the McVittie solution cannot represent a physically realistic inhomogeneity is, we think, best shown by a glance at Figure \ref{figure10} and equation (\ref{hrelation}). By construction, no fluid streamlines can cross $\mathcal{H}$ and so the black and white hole horizons are present here by way of mathematical extensions, not physical processes. There is no easy fix for this situation, within the context of McVittie's approach, since a routine calculation shows that (\ref{metric1}) is a perfect fluid if and only if $dm/dt=0$. Nonetheless, the present analysis shows that a rather routine looking spacetime, like (\ref{metric1}), can in fact harbor a rather exotic interior. Moreover, the present analysis suggests that the very notion of an inhomogeneity in cosmology may go beyond the concept of inhomogeneity in elementary physical variables.

What we have done here can be expanded in a number of ways. First, of course, one could relax condition (\ref{Hspecial}) and consider a wider class of possibilities. In all cases one would find that if the vacuum boundary ($\mathcal{H}$) contains a bifurcate two - sphere, then this bifurcate two - sphere is also part of the McVittie solution itself. This geometric behaviour can be traced to McVittie's no - flux condition which preserves the integrity of $\mathcal{H}$. One can reasonably expect that the integrity of $\mathcal{H}$ is destroyed by any flux through it. To conclude, we believe that the McVittie solution is an instructive idealization very much like the Kruskal - Szekeres extension.

\begin{acknowledgments}
This work was supported in part by a grant from the Natural Sciences and Engineering Research Council of Canada (to KL) and a Carmichael Fellowship (to MA). We have several people to thank for discussions which have improved our paper. In the early stages we benefited from discussions with David Garfinkle and Matt Visser. As the project developed, we had extensive discussions with Damien Martin, Matthew Kleban and Brien Nolan. Of course, we do not mean to suggest that these people necessarily agree or disagree with what we have concluded here.
\end{acknowledgments}

\appendix

\section{Ingoing null geodesics}

The arguments presented here\footnote{Some of the conclusions in this Appendix (and we expect by now all) have also been obtained by Brien Nolan.} do not rely on a specific form for $H$. We are concerned only with ingoing radial null geodesics. It is convenient to write the associated null geodesic equation in the form
\begin{equation}\label{affinet}
    \frac{d^2t}{d \lambda^2}=-\left(\frac{H(1-\frac{m}{R})}{\sqrt{1-\frac{2m}{R}}}-\frac{2m}{R^2}\right)\left(\frac{dt}{d \lambda}\right)^2.
\end{equation}
\subsection{{\large $\eta$}$_{1}$}
Along {\large $\eta$}$_{1}$, $f<0$ and defining
\begin{equation}\label{h}
    h(R) \equiv \left(\frac{3m}{R}-1\right)\frac{1}{R}
\end{equation}
we have
\begin{equation}\label{n1h}
    \frac{d^2t}{d \lambda^2}<h(R)\left(\frac{dt}{d \lambda}\right)^2.
\end{equation}
Now along {\large $\eta$}$_{1}$, $R$ is strictly increasing to $R_{+}>3m$ and so $h(R)$ eventually becomes negative. Define some fiducial $R_{0}$ so that
\begin{equation}\label{fiducial}
    h(R_{0}) =-\alpha
\end{equation}
where $0<\alpha<1/12m$. Now (\ref{fiducial}) gives
\begin{equation}\label{fiducials}
    R_{0}^{\pm}=\frac{1\pm\sqrt{1-12 \alpha m}}{2 \alpha}
\end{equation}
and since we can always choose $\alpha$ sufficiently small (but not zero) so that $R_{0}^{-}<R_{+}<R_{0}^{+}$ we eventually have
\begin{equation}\label{limitn1}
    \frac{d^2t}{d \lambda^2}<-\alpha \left(\frac{dt}{d \lambda}\right)^2
\end{equation}
along {\large $\eta$}$_{1}$ for $\lambda>$ some $\lambda_{*}$. Now write
\begin{equation}\label{Tdef}
    \frac{dt}{d \lambda}=T
\end{equation}
so that from (\ref{affinet}) we have
\begin{equation}\label{first}
    \frac{dT}{d \lambda}<-\alpha T^2.
\end{equation}
Integrating (\ref{first}) we have
\begin{equation}\label{second}
    T<\frac{1}{\alpha(\lambda-\lambda_{*})-1/T_{*}}.
\end{equation}
Finally, integration of (\ref{second}) gives
\begin{equation}\label{third}
    t-t_{*}<\frac{1}{\alpha}\ln(T_{*} \alpha (\lambda-\lambda_{*})+1).
\end{equation}
From (\ref{third}) it follows that $t \nrightarrow \infty$ for finite $\lambda$ and so {\large $\eta$}$_{1}$ is geodesically complete.
\subsection{$(b,i^{+})$}

We consider the ingoing geodesics which reach $\mathcal{H}$ within the range $(b,i^{+})$. Sufficiently close to $\mathcal{H}$, $R$ is strictly decreasing and $f \geq 0$ with equality holding only on $\mathcal{H}$. We now have
\begin{equation}\label{n3h}
    \frac{d^2t}{d \lambda^2}\geq h(R)\left(\frac{dt}{d \lambda}\right)^2
\end{equation}
where we continue to use (\ref{h}). Now since $R$ approaches $R_{-}<3m$, $h(R)$ is strictly positive and increasing. Again, for some fiducial $R_{0}$, but now $>R_{-}$, write $h(R_{0})= \alpha>0$. We now have
\begin{equation}\label{limitn3}
    \frac{d^2t}{d \lambda^2}\geq \alpha \left(\frac{dt}{d \lambda}\right)^2
\end{equation}
for $\lambda>$ some $\lambda_{*}$. Now integrating in parallel to the details given in the above case we arrive at
\begin{equation}\label{third3}
    t-t_{*}\geq \frac{1}{\alpha}\ln\left(\frac{1}{T_{*} \alpha (\lambda_{*}-\lambda)+1}\right).
\end{equation}
From (\ref{third3}) it follows that $t \rightarrow \infty$ for finite $\lambda$ and so ingoing geodesics that reach $\mathcal{H}$ in the range $(b,i^{+})$ are geodesically incomplete. This observation is the principle contribution given in KKM.
\subsection{$(^{-}i^{0},b]$}

Finally, we consider the ingoing geodesics which reach $\mathcal{H}$ within the range $(^{-}i^{0},b]$. This includes {\large $\eta$}$_{2}$. Now $R$ is strictly increasing and $f \leq 0$ with equality holding only on $\mathcal{H}$. We now have
\begin{equation}\label{n2h}
    \frac{d^2t}{d \lambda^2}\leq h(R)\left(\frac{dt}{d \lambda}\right)^2
\end{equation}
where we continue to use (\ref{h}). Now since $R$ approaches $R_{-}<3m$, $h(R)$ is strictly positive and decreasing. Again, for some fiducial $R_{0} <R_{-}$, write $h(R_{0})= \alpha>0$. We now have
\begin{equation}\label{limitn2}
    \frac{d^2t}{d \lambda^2}\leq \alpha \left(\frac{dt}{d \lambda}\right)^2
\end{equation}
for $\lambda>$ some $\lambda_{*}$. Now integrating in parallel to the details given in the above cases we arrive at
\begin{equation}\label{third2}
    t-t_{*}\leq \frac{1}{\alpha}\ln\left(\frac{1}{T_{*} \alpha (\lambda_{*}-\lambda)+1}\right).
\end{equation}
From (\ref{third2}) it follows that $t \rightarrow \infty$ for finite $\lambda$ and so ingoing geodesics that reach $\mathcal{H}$ in the range $(^{-}i^{0},b]$ are also geodesically incomplete.

\section{Bifurcation two - spheres}

A bifurcation two - sphere is usually discussed in terms of a vanishing time-translational Killing vector (e.g. \cite{poisson}). Here we define a bifurcation two - sphere in terms of the simultaneous vanishing of both the ingoing and outgoing radial null geodesic expansions. First, for clarity, let us review the situation in the Schwarzschild vacuum. As shown in \cite{lake}, the Kruskal - Szekeres metric can be given as
\begin{equation}\label{basemetric}
ds^2=(2M)^2 d\tilde{s}^2
\end{equation}
with
\begin{equation}\label{ks2}
d\tilde{s}^2=\frac{-4}{(1+\mathcal{L})e^{1+\mathcal{L}}}dudv+(1+\mathcal{L})^2d\Omega^2_{2}
\end{equation}
where
\begin{equation}\label{L}
    \mathcal{L} \equiv \mathcal{L}(-\frac{u v}{e})
\end{equation}
and $\mathcal{L}$ is the Lambert W function \cite{lambert}.
 Trajectories
with tangents $k^{\alpha}=e^{\mathcal{L}}(1+\mathcal{L}) \delta^{\alpha}_{v}$ (constant $u=u_{0}, \theta$ and $\phi$) are radial null geodesics given by
\begin{equation}\label{u0}
    v(\lambda)=\lambda e^{-\frac{u_{0} \lambda}{e}}
\end{equation}
where $\lambda$ is an affine parameter and we note the expansion
\begin{equation}\label{expK}
    \nabla_{\alpha}k^{\alpha}=\frac{-2u_{0}}{e(1+\mathcal{L})}.
\end{equation}
Trajectories
with tangents $l^{\alpha}=e^{\mathcal{L}}(1+\mathcal{L}) \delta^{\alpha}_{u}$ (constant $v=v_{0}, \theta$ and $\phi$) are radial null geodesics given by
\begin{equation}\label{v0}
    u(\lambda)=\lambda e^{-\frac{v_{0} \lambda}{e}}
\end{equation}
and we now note the expansion
\begin{equation}\label{expm}
    \nabla_{\alpha}l^{\alpha}=\frac{-2v_{0}}{e(1+\mathcal{L})}.
\end{equation}
On the horizons $u=0$ and $v=0$ then $v$ and $u$ are affine parameters. The bifurcation two - sphere is given by $u = v = 0$ and it is uniquely characterized by $\nabla_{\alpha}k^{\alpha}=\nabla_{\alpha}l^{\alpha}=0$.

\bigskip

The McVittie solution under consideration in this paper reduces \cite{schds} to the Schwarzschild - de Sitter spacetime on $\mathcal{H}$ and so we need a more general construction. We follow \cite{ballik}. It is shown there that for all static metrics
\begin{equation}\label{fform}
    ds^2=-fd t^2+\frac{d r^2}{f}+r^2d\Omega^2_{2},
\end{equation}
where $f$ is a polynomial with simple root(s),
\begin{equation}\label{froot}
    f(r)=(r-a)h(r)
\end{equation}
where $h(a)\neq 0$, one can construct regular extensions about $r=a$ via the transformations
\begin{equation}\label{rtrans}
    uv=\pm(r-a)\exp(\int 2 \kappa \frac{k(r)}{h(r)}dr+\mathcal{E}),
\end{equation}
where the sign depends on how we choose to orientate the $u-v$ axis, $\mathcal{E}$ is a constant and
\begin{equation}\label{ttrans}
    \Big|\frac{v}{u}\Big|=\exp(2 \kappa t)
\end{equation}
where $\kappa$ is the surface gravity given by
\begin{equation}\label{surf}
    \kappa \equiv \frac{1}{2}\frac{df}{dr}\Big|_{a} \neq 0.
\end{equation}
Note that according to (\ref{rtrans}) $r=r(uv)$ and $r^{'}|_{a}\neq 0, ^{'}\equiv d/duv$.
Under these transformations the Killing vector $\eta^{\alpha}=\delta^{\alpha}_{t}$ becomes $\eta^{\alpha}=(u,-v,0,0)$ and one recovers the usual definition of the bifurcation two - sphere at $u=v=0$. Note that the specified construction can always be done. However, about a distinct root, say $r= b \neq a$, a new chart must be constructed about $r=b$.

To calculate null geodesic expansions note that the metric takes the form
\begin{equation}\label{kruskal}
    ds^2=K(r)dudv+r^2d\Omega^2_{2}
\end{equation}
where
\begin{equation}\label{K}
    K(r)\equiv\pm \frac{a h(r)}{\kappa^2} \exp(-2 \kappa \int \frac{k(r)}{h(r)}dr).
\end{equation}
The integration constant has been absorbed into the factor $a$ and again the choice of sign determines the orientation of the $u-v$ axis.
Trajectories with tangents $k^{\alpha}=\frac{ \delta^{\alpha}_{v}}{K(r)}$ (constant $u=u_{0}, \theta$ and $\phi$) are radial null geodesics with expansions
\begin{equation}\label{expKg}
    \nabla_{\alpha}k^{\alpha}=\frac{u_{0}}{K(r)r}r^{'},
\end{equation}
and trajectories
with tangents $l^{\alpha}=\frac{ \delta^{\alpha}_{u}}{K(r)}$ (constant $v=v_{0}, \theta$ and $\phi$) are radial null geodesics with expansions
\begin{equation}\label{expmg}
    \nabla_{\alpha}l^{\alpha}=\frac{v_{0}}{K(r)r}r^{'}.
\end{equation}
The bifurcation two - sphere associated with any non - degenerate horizon at $r=a$ is therefore characterized by the simultaneous vanishing of both the ingoing and outgoing radial null geodesic expansions; $\nabla_{\alpha}k^{\alpha}=\nabla_{\alpha}l^{\alpha}=0$.

\section{The degenerate case $27m^2H_{0}^2=1$}

\begin{figure}[ht]
\epsfig{file=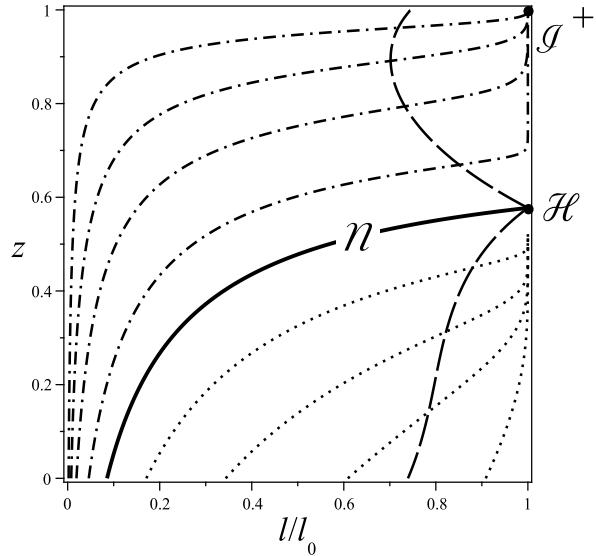,height=3in,width=3.5in,angle=0}
\caption{\label{figure13} As in Figure \ref{figure5} but for $l_{0}=1/H_{0}m=3\sqrt{3}$ (again we take $H_{0}=1/3$). The limiting geodesics {\large $\eta$}$_{1}$ and {\large $\eta$}$_{2}$ now coincide and give {\large $\eta$}. There is no locus $f=0$. The dashed curves connecting $z=0$ and $\mathcal{H}$ and $\mathcal{H}$ with $z=1$ indicate $d^2z/dl^2=0$.}
\end{figure}

Since $\dot{H}<0$, $f_{0} \geq f$. In the degenerate case $27m^2H_{0}^2=1$ and so
\begin{equation}\label{degen}
    f \leq -\frac{(3m-R)^2(6m+R)}{27m^2R}.
\end{equation}
Throughout the associated McVittie solution $f<0$ and $f=0$ only at the horizon $\mathcal{H}$ where $R=3m$. As a result, all ingoing and outgoing radial null geodesics have $dR/dt>0$ and there can be no bifurcation two - sphere. All ingoing geodesics that reach $\mathcal{H}$ do so without crossing $f=0$ first and as in the non - degenerate case we find that these ingoing geodesics are incomplete. The ingoing radial null geodesics in the $z - l$ plane are shown in Figure \ref{figure13}.

\begin{figure*}[ht]
\epsfig{file=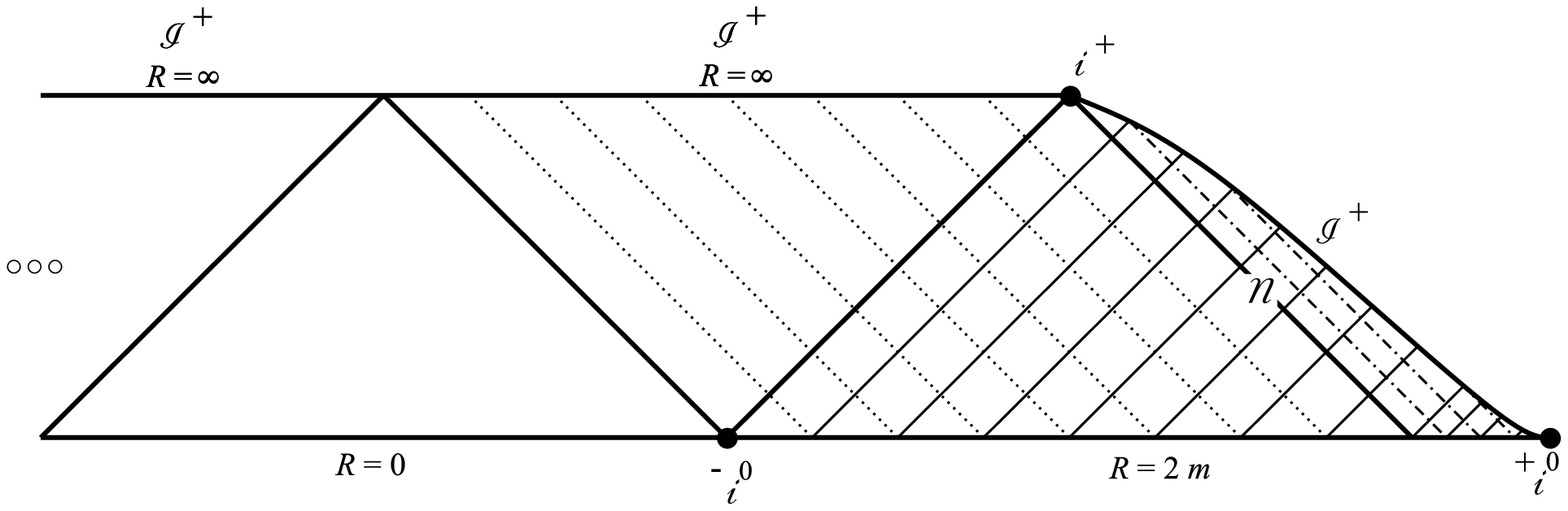,height=3in,width=6in,angle=0}
\caption{\label{figure14} As in Figure \ref{figure12} but now for the degenerate case.}
\end{figure*}

A possible extension of the associated McVittie solution is shown in Figure \ref{figure14} where we have used the mapping function
\begin{equation}\label{mapdegen}
t=A(1-x){\tan\left((1-x)\frac{\pi}{4}\right)}
\end{equation}
with $A \sim 1.9$ and a degenerate Schwarzschild - de Sitter ``interior".

\section{$H_{0}=0$}

\begin{figure}[ht]
\epsfig{file=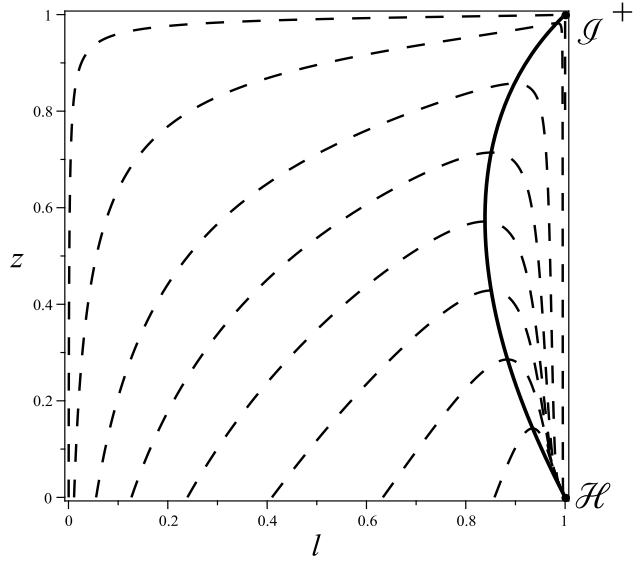,height=3in,width=3.5in,angle=0}
\caption{\label{figure15}As in Figure \ref{figure5} but for $H_{0}=0$. Note that $\mathcal{H}$ is now given by ($R= 2m ,t = \infty$).}
\end{figure}

\begin{figure}[ht]
\epsfig{file=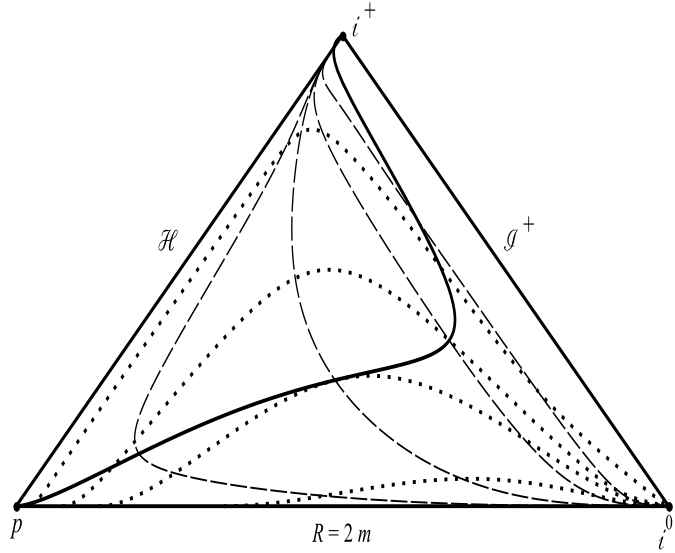,height=3in,width=3.5in,angle=0}
\caption{\label{figure16}Conformal representation of surfaces of constant $R$ (dashed) and constant $t$ (dots) for the case $H_{0}=0$. The locus $f=0$ from $p$ to $i^{+}$ is shown (solid).}
\end{figure}

The case $H_{0}=0$ proceeds in a fundamentally different way than the case $H_{0}>0$. To see this, consider the usual background of ``dust" so that $H=2/3t$. Defining $x=R/m$ and $T=t/m$ the locus $f=0$ takes the form
\begin{equation}\label{locuszero}
    T=\frac{2x^{3/2}}{3 \sqrt{x-2}}
\end{equation}
and so surfaces of constant $t$ never intersect the locus for $t<2 \sqrt{3}m$, intersect it once for $t=2 \sqrt{3}m$ and intersect it twice for $t>2 \sqrt{3}m$. Moreover, every surface of constant $R$, for $2m < R < \infty$, crosses the locus once and at finite $t$. This last point shows us that eventually all ingoing radial null geodesics have $f>0$ and so eventually $dR/dt<0$. There is no bifurcation two - sphere in this McVittie spacetime. To integrate the radial null geodesics we continue to use (\ref{z}) but replace (\ref{l}) with
\begin{equation}\label{lnolambda}
    l=\frac{1}{1+mH}.
\end{equation}
The radial ingoing null geodesic equations now take the form
\begin{equation}\label{geonolambda}
    \frac{dz}{dl}=\frac{(1-z^2)^2}{6(1-l^2)}\left(\frac{2(1-l)}{l(1-z^2)} - z \right).
\end{equation}
The integrations are shown in Figure \ref{figure15}.
Using the same procedures as before, the surfaces of constant $R$ and constant $t$ are shown in Figure \ref{figure16}. We find that $\mathcal{H}$ is at finite affine distance along ingoing null geodesics but $\mathcal{I}^{+}$ is at infinite affine distance along outgoing null geodesics (this spacetime is asymptotically flat).

\bigskip

\begin{figure*}[ht]
\epsfig{file=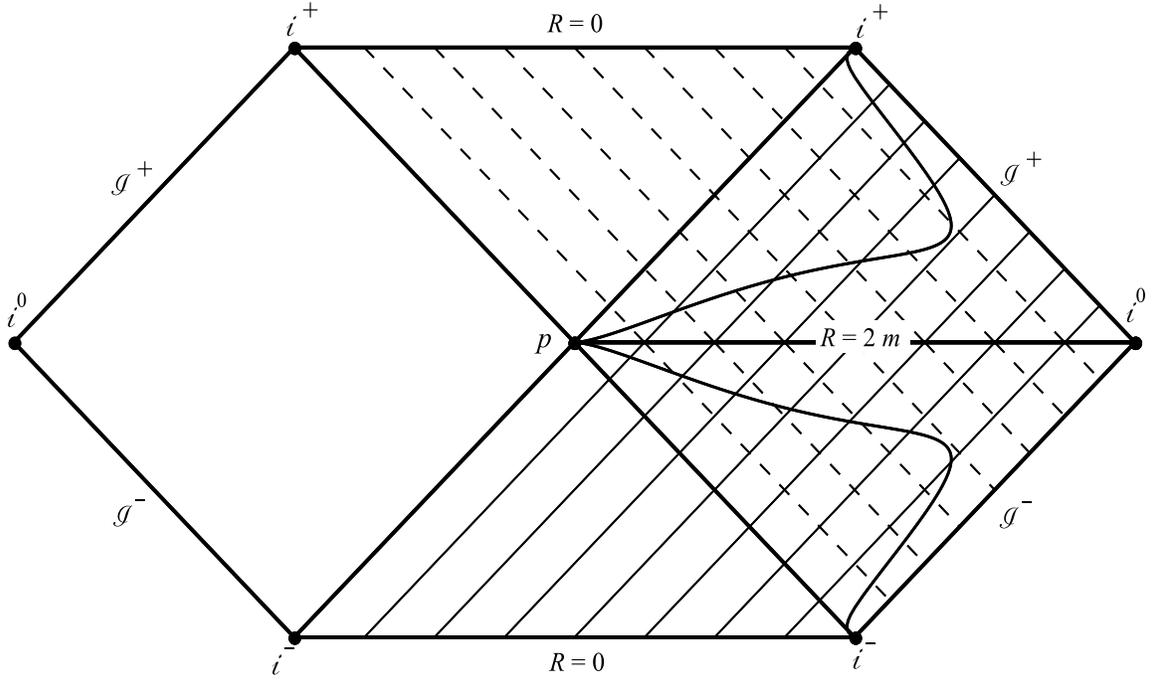,height=4in,width=6in,angle=0}
\caption{\label{figure17}An extension of the McVittie spacetime for $H_{0}=0$ which is null geodesically complete. Note that all trajectories terminate on the spacelike singularity $R=2m$. The loci $f=0$ are shown.}
\end{figure*}

The conformal diagram and a possible extension onto the Kruskal - Szekeres manifold is shown in Figure \ref{figure17}. There is now a black hole horizon to the future of the singularity and a white hole horizon in the past. These now form a wedge, not a straight line.

\bigskip

Now KKM argue, their considerations motivated by quantum gravity arguments, that the invariant
\begin{equation}\label{invariant}
    \Delta \equiv \nabla_{\iota}\nabla_{\epsilon}R_{\alpha \beta \gamma \delta}\nabla^{\iota}\nabla^{\epsilon}R^{\alpha \beta \gamma \delta}
\end{equation}
 diverges on the horizon. It is difficult to see how this would come about since at the horizon, using conditions (\ref{Hcon}), with $H_{0} \geq 0$, the metric tensor along with all first and second order (partial derivatives) are continuous. Using GRTensor II \cite{grt}, and assuming $H$ is $\in C^3$, we find
\begin{equation}\label{newinvariant}
    \lim_{t\rightarrow\infty}\Delta=\frac{1440m^2}{R^{12}}( H_{0}^2R^3(11H_{0}^2R^3-24R+50m)+14R^2-60Rm+65m^2)
\end{equation}
where evaluation along $\eta_{-}$ is understood. We see no divergence for $R > 0$.


\begin{thebibliography}{}
\bibitem{notation} We use geometrical units and a signature of +2 throughout.
Functional dependence is designated only on the first
appearance of a function.
\bibitem{majd}The choice of the shape of the singularity $R=2m$ as a straight horizontal line is arbitrary, as is the choice of the time transformation function (\ref{conformalT}). Any spacelike curve (i.e. $-1<slope<1$ ) with a finite length, and any function, that is monotonically increasing, and spans $ 0<t<\infty $ in a finite interval of the new variable, could be used to construct the conformal diagram.
\bibitem{schds} With conditions (\ref{Hcon}) the metric tensor and all derivatives to order 2 are continuous accross $\mathcal{H}$ and so matching accross $\mathcal{H}$ is complete under any formulation of matching conditions.
\end{thebibliography}
\end{document}